\documentclass[letterpaper]{article}
\usepackage{aaai2027}  
\nocopyright           
\usepackage[hyphens]{url}
\usepackage{graphicx}
\usepackage{natbib}
\usepackage{caption}
\usepackage{amsmath,amssymb}
\usepackage{booktabs}

\AtBeginDocument{\ifdefined\Urlmuskip\Urlmuskip=0mu plus 2mu\relax\fi}

\begin{document}

\title{Toward Measuring AI's Effects on Skill Formation:\\ The Stock--Formation Gap}
\author{Aysa X. Fan}
\affiliations{University of Illinois Urbana-Champaign, Urbana, IL, USA}

\maketitle
\pagestyle{plain}

\begin{abstract}
Large-scale AI deployment data and controlled learning experiments characterize different consequences of the same technology. Deployment telemetry shows that AI use is concentrated in skilled work and frequently supports immediate task performance. It observes tasks, interaction patterns, and outputs, however, not whether users become more capable of performing those tasks independently. Controlled studies measure independent capability more directly, but only in narrower populations and settings, with outcomes that vary substantially by interaction design. We formulate this discrepancy as a stock--formation measurement gap: current systems observe the use of existing expertise more readily than the formation of future expertise. Because formation has historically been society's recovery mechanism through technological change, the gap matters well beyond any single classroom. We synthesize the experimental and observational evidence by identification strength, use public deployment data as a descriptive illustration of the gap, and identify the missing bridge between interaction traces and unassisted retention and transfer. We then propose a research program that links consented usage records to independent assessments while experimentally varying whether AI supplies answers, hints, feedback, or evaluation. The claim is not that AI has been shown to erode skill formation at population scale. It is that existing measurement cannot determine whether it does, and that this question is both measurable and designable.
\end{abstract}

\section{The Measurement Puzzle}\label{sec:puzzle}

Two bodies of evidence describe how people use AI, and they are often read as if they disagree. Deployment telemetry, most extensively the Anthropic Economic Index built from over four million assistant conversations, shows AI use concentrated among skilled professionals, dominated by augmentation-style collaboration (roughly 57\% augmentation versus 43\% automation), and rewarded by experience \citep{handa2025economic,anthropic2026aeiv5}. Productivity gains scale with the schooling a task requires \citep{appel2026}, and field experiments find real performance improvements within the technology's capability frontier, alongside degraded performance beyond it \citep{brynjolfsson2025generative,dell2023navigating}. Controlled learning experiments, meanwhile, repeatedly find that AI assistance can raise immediate task performance while leaving subsequent unassisted capability flat or reduced, with the outcome varying substantially by how the interaction is designed \citep{fan2024,bastani2025,contractor2026learning}.

These findings are not in conflict, because they concern different quantities. Telemetry observes tasks, interaction patterns, and immediate outputs. The experiments measure whether a person can later perform without the tool. The first quantity is about the use of existing capability; the second is about the formation of new capability. Reading either literature as an answer to the other's question produces most of the apparent disagreement in current debates about AI and skills.

This paper formulates the discrepancy as a measurement problem, which we call the stock--formation gap: the measurement infrastructure now watching AI adoption observes the deployment of the existing stock of human capital far more readily than the formation of the next generation of it. The central claim is deliberately narrow. The concern is not that current evidence proves AI is eroding learning at scale. The concern is that the dominant measurement infrastructure can register improvements in assisted output while remaining unable to observe changes in independent capability. Controlled studies can measure capability formation but not at deployment scale; deployment telemetry has the scale but does not observe formation; the open question sits where neither instrument reaches.

Why the question deserves attention is a matter of history. Across two centuries of technological change, society's principal \emph{recovery mechanism} has been education: when technology displaced workers, the next generation adapted by forming the capabilities the new economy demanded. Inequality dynamics track whether the supply of educated workers keeps pace with technological demand \citep{tinbergen1975,goldin2008}, and skills account for most of the variation in workers' earnings \citep{frey2019}. That history rests on an assumption so embedded it is rarely stated: however disruptive a technology, the formation process itself stays intact. Previous technologies automated lower-order cognitive work, and education adapted by moving instruction toward the higher-order work machines could not reach. General-purpose AI is among the first technologies to which learners can, and observably do, hand much of the higher-order work itself; in the largest student dataset, Creating and Analyzing are the most delegated levels (Section~\ref{sec:evidence-t3}). Whether that delegation erodes formation is an open empirical question; the trials suggest the answer depends on the allocation of cognitive work rather than on the technology as such. The survival of the recovery mechanism should be a measured quantity, not an article of faith.

Exposure, meanwhile, is already at scale. Coursework alone accounts for roughly one in eight conversations on a major assistant platform, and directive interaction, in which the user delegates a task and receives an output, is the platform's most common mode overall (32.6\%) \citep{anthropic2026aeiv5}. A parallel economics literature has begun to model the workplace version of the same concern, in which automation of entry-level tasks removes the channel through which juniors acquire tacit skill \citep{ide2025intergenerational,garicano2025training,afrouzi2026career}, with early empirical counterparts in junior hiring declines at AI-adopting firms \citep{hosseini2025seniority,brynjolfsson2026canaries}. That literature is production-side; the measurement side, to our knowledge, has not been formulated.

The ingredients of the stock--formation formulation are individually well established: the performance-learning distinction is classical \citep{soderstrom2015learning}, cognitive offloading and over-reliance have substantial literatures, and trace-based skill modeling is routine on dedicated tutoring platforms (Section~\ref{sec:misses-bridge}). What we take to be new is the connection: the population composition of who is visible in deployment data, the mismatch between where scale lives and where outcomes are measured, and a concrete measurement bridge for general-purpose assistants. The stock--formation gap is not a renaming of performance versus learning; it is a claim about which of those quantities existing infrastructure can observe, at what scale, and for whom. This is deliberately a position and agenda paper: it contributes a formulation, an evidence synthesis, and a research program rather than new experimental results; its one original analysis is descriptive by design, and our own failed pilots appear as design lessons (Section~\ref{sec:program-lessons}). The studies the agenda specifies are its intended sequel, not its content.

The paper proceeds as follows. Section~\ref{sec:framework} defines the framework's seven quantities. Section~\ref{sec:evidence} synthesizes the experimental and observational evidence, organized by identification strength with respect to the variable that appears to matter most: who performs the cognitive work, which Section~\ref{sec:framework} separates into the condition the system assigns and the allocation the learner actually realizes. Section~\ref{sec:misses} maps what the existing evidence base measures and what it does not, using one descriptive analysis of public deployment data as an illustration. Section~\ref{sec:program} develops the research program that would close the gap; it is the point of the paper. Section~\ref{sec:implications} states what results from that program would mean, and the limits of the present synthesis.

\section{The Stock--Formation Framework}\label{sec:framework}

Seven quantities organize everything that follows.

\begin{table*}[t]
\centering
\caption{The stock--formation gap, stated as what each data source observes and what it can support.}
\label{tab:stock-formation}
\begin{tabular}{p{2.5cm}p{4.7cm}p{4.7cm}}
\toprule
 & \textbf{Deployment telemetry} & \textbf{Controlled learning studies} \\
\midrule
Population & Mostly stock ($S$): skilled workers & Mostly formation ($F$): students, novices \\
Scale & Millions of conversations & Tens to $\approx$1{,}000 participants \\
Observes & Task, mode labels (noisy correlates of $A$), output $Y$ & Unassisted capability change $\Delta K$ \\
Does not observe & $\Delta K$; who is learning what & Population-scale usage; long-run field behavior \\
Can support & Descriptions of use of existing expertise & Causal effects of assigned conditions $Z$ in specific settings \\
\bottomrule
\end{tabular}
\end{table*}

Let $S$ denote the \emph{stock} state: a person whose capability in a given domain is largely formed and who uses AI to deploy it. Let $F$ denote the \emph{formation} state: a person whose capability in that domain is still forming. These are domain- and time-relative positions of a person--domain pair, not types of people: a professional learning an unfamiliar library is in formation for that domain \citep{shen2026skill}, and a student may be deploying already-formed expertise elsewhere; the Bridge and Mixed entries in Table~\ref{tab:gap-matrix} mark studies that straddle the two, and a continuum refinement is natural. We still speak of stock and formation \emph{populations} as compositional shorthand: coursework contexts are formation-heavy and professional telemetry is stock-heavy on average, and composition is what determines what deployment data can be taken to show.

For any human--AI interaction, let $Z$ denote the \emph{assigned condition}: the interface, defaults, or configuration under which assistance is delivered, a spectrum running from answer delivery through hints and feedback to evaluation of the person's own attempt. Let $A$ denote the \emph{realized allocation of cognitive work}: which parts of the task's cognitive work the person actually performs in use. The two are not equivalent: a hints interface can be used passively, and an answer-delivery interface can prompt active verification, so $Z$ shapes but does not determine $A$. Nor is $A$ a single answer-delivery-to-evaluation scale; it is best treated as a profile over component functions, for instance planning, generation, monitoring, verification, and correction, and we intend the profile reading wherever measurement is at issue. Let $Y$ denote the immediate output (the essay drafted, the code produced, the problem answered). Finally, let $K_0$ and $K_1$ denote \emph{unassisted} capability at baseline and after a period of use, measured with the tool removed and including retention over time and transfer to novel problems, with $\Delta K = K_1 - K_0$.

The $Z$--$A$ distinction parallels a recent reframing in learning analytics of learner agency as the allocation of decision authority across learners, systems, and institutions, where a recurring obstacle is that agency is inferred from behavioral proxies rather than measured directly \citep{borchers2026agency}. The two allocations can vary independently: a learner who retains full decision authority can still delegate the thinking itself, and a system that architects every choice can still leave the cognitive work with the learner. $A$ tracks the cognitive-work allocation, and it inherits the same behavioral-proxy obstacle, which is why the program in Section~\ref{sec:program} treats measuring $A$ from traces as a first-class validation problem rather than an assumption.

The distinction between $Y$ and $\Delta K$ is the learning-science distinction between performance and learning: performance is observable execution during acquisition, while learning is the durable change that survives a delay and a change of conditions, and the two are empirically dissociable in both directions \citep{soderstrom2015learning}. A second regularity gives the allocation variable its learning-science meaning: conditions that make practice feel harder, the retrieval, monitoring, and error correction that register as struggle, often produce more durable learning than conditions that make performance smooth, the pattern known as desirable difficulties \citep{bjork2011}. On that account, the effortful components of a task are not overhead around learning; they are part of the mechanism of formation, and an assistant that absorbs them by default would absorb formation opportunities along with the friction. That is the mechanistic reason the allocation of cognitive work is the natural moderator, and why effects on $\Delta K$ cannot be read off improvements in $Y$. These distinctions do all the work in what follows, so we state them once and rely on them throughout.

In these terms: deployment telemetry observes task, interaction-mode labels that are at best noisy correlates of $A$, and $Y$, mostly in $S$; it observes neither $K_0$ nor $K_1$. Randomized experiments assign $Z$ and estimate its effect on $\Delta K$, mostly in $F$, at small scale; because assignment operates on $Z$ while the hypothesized mechanism runs through $A$, claims about $A$ itself require mediation or observational analysis even inside a trial. The quantity that matters for how a society adapts to this technology is the effect of sustained use, under the conditions people actually experience and the allocations they actually realize, on $\Delta K$ in $F$ at population scale. No measurement system we are aware of estimates it (Table~\ref{tab:stock-formation}).

\section{What the Evidence Establishes}\label{sec:evidence}

We synthesize the evidence in three tiers by identification strength with respect to the assigned condition $Z$ and the realized allocation $A$. The tiers matter because the literatures use different vocabularies for these variables (unscaffolded versus scaffolded arms in trials are values of $Z$; directive versus learning-oriented labels in telemetry and externalization versus internalization in the education literature are approximate descriptions of $A$), and because the strength of causal claims differs sharply across them.

\subsection{Tier 1: randomized assignment of the interface ($Z$)}\label{sec:evidence-t1}

Three experiments randomize the assigned condition itself, and they carry the strongest conclusions. \citet{bastani2025}, in a randomized trial with approximately one thousand high-school mathematics students, found that access to a base GPT-4 tutor raised performance on assisted practice problems by 48\% while lowering subsequent unassisted exam scores by 17\% relative to no-AI control. A second arm deployed the same model behind pedagogical guardrails that offered hints rather than solutions: assisted practice improved even more (127\%), and the scaffolded condition did not show a statistically detectable deficit relative to control, suggesting, but not establishing, that interaction design can mitigate the effect. Because the intervention was a prompting-and-guardrail layer rather than a new model, the design lever it isolates is available to any deployer.

\citet{alsaiari2026directive}, in a semester-long trial ($n = 329$), randomized the type of AI feedback students received. Directive feedback produced roughly three times higher odds of revision than purely metacognitive prompts ($\mathrm{OR} = 2.93$, $p = .009$), and hybrid feedback outperformed both (27.5\% versus 12.1\% revision). The finding cuts against a simple ``more learner effort is always better'' reading: novices given only reflective prompts often failed to convert them into action, so learner-side load can be overloaded as well as offloaded. \citet{kumar2026empathy}, in a pre-registered trial ($n = 968$), tested the inverse allocation directly: the participant practiced empathic communication and the AI evaluated and guided, producing a 0.98~SD improvement across most measured dimensions. Together, Tier~1 shows that within these settings the assigned condition $Z$ causally affects $\Delta K$, in both directions. The natural reading, that $Z$ works by shifting the realized allocation $A$, is plausible but not directly established: these designs randomized interfaces and feedback types, not cognitive behavior, and $A$ was at most partially observed.

\subsection{Tier 2: randomized access, observational usage ($A$ observed post hoc)}\label{sec:evidence-t2}

A second set of experiments randomizes access to AI but observes how participants use it. \citet{fan2024} ($n = 117$) found ChatGPT support significantly improved essay scores while producing no detectable difference in knowledge gain, transfer, or intrinsic motivation; trace analysis showed reduced self-monitoring in the AI condition, a pattern the authors call metacognitive laziness. \citet{shen2026skill} found professional software engineers learning a new library with AI scored 17\% lower on a subsequent quiz ($d = 0.738$) with no significant completion-time gain; post hoc coding of screen recordings associated preserved learning with engagement-style patterns and losses with delegation-style patterns, though pattern membership was self-selected. \citet{contractor2026learning} ($n = 211$, proctored) found the opposite headline: assistant access raised unaided test scores by 0.27~SD, with most of the gain persisting at one week; within the treatment arm, gains concentrated among participants who used the assistant to explain concepts, while those who used it to generate text lost their gains once it was removed, again a post hoc contrast. \citet{liu2026persistence}, across trials totaling 1{,}222 participants, found assistance improved immediate performance but harmed subsequent unassisted performance ($d \approx -0.42$), concentrated among participants who requested solutions rather than hints. \citet{barcaui2025crutch} ($n = 120$) found unrestricted chatbot users scoring eleven percentage points below traditional-study controls on a surprise test 45 days later.

Tier~2 establishes that access effects on $\Delta K$ are real and heterogeneous, and it repeatedly finds outcomes tracking the observed allocation $A$: how much of the cognitive work the person retained. But $A$ is self-selected in this tier, so selection (stronger or more motivated learners choosing explanation-seeking use) cannot be excluded.

\subsection{Tier 3: observational telemetry and platform data}\label{sec:evidence-t3}

The deployment-scale evidence is observational, and outside of two instrumented platforms it contains no learning outcomes. The Anthropic Student Report ($574{,}740$ student conversations) classifies nearly half of student interactions ($\sim$47\%) as ``Direct'': instructions given, output received, minimal engagement; students also delegate the upper levels of the cognitive-process hierarchy (Creating, 39.8\%; Analyzing, 30.2\%) far more than lower levels \citep{anthropic2025student,anderson2001taxonomy}. Some Direct interactions may instantiate the answer-delivery mechanism manipulated in the unscaffolded trial arms, but this correspondence has not yet been validated, and the report itself notes that usage labels cannot establish learning impact. The Educator Report finds nearly half of educator assessment interactions automate grading \citep{bent2025anthropic}, which matters here for a measurement reason developed in Section~\ref{sec:misses}. On platforms with built-in assessment, more can be seen: across 3.2 million interactions on an instrumented adaptive-learning platform, study time on AI-susceptible problems fell after ChatGPT's release while proctored retention performance declined and unproctored performance rose \citep{rismanchian2026faster}; and in competitive programming telemetry, an AI-reliant practice signature predicted smaller skill gains for substitute-style but not complement-style users \citep{yao2026scaffold}.

At the synthesis level, a meta-analysis reports large average performance gains from ChatGPT use ($g = 0.7$) \citep{deng2025}, but reviews that separate outcome types read the same literature as ``short-term boost, uncertain transfer,'' with only a small minority of studies providing strong causal evidence \citep{fesler2026evidence,yan2025}. A Bayesian living meta-analysis in mathematics education estimates a more modest pooled effect ($g = 0.42$, 95\% CrI $[0.13, 0.72]$) and flags substantial publication bias \citep{strohmaier2026llama}. A systematic review of fifty-six studies finds the best learning outcomes under high-human-control designs, and also that only 21\% of studies met minimum power requirements \citep{liang2026systematic}; a \emph{Nature Human Behaviour} commentary warns of an ``AI-induced performance illusion'' when assisted performance is mistaken for learning \citep{yan2024}.

\subsection{What can and cannot be concluded}\label{sec:evidence-conclusions}

Three statements summarize the evidence honestly. First, in randomized settings, AI assistance reliably improves immediate assisted performance; this is not in dispute. Second, effects on unassisted capability vary from clearly negative to clearly positive, and the variation consistently tracks either the assigned condition $Z$ (Tier~1) or the observed allocation $A$ (Tier~2); $Z$ was randomized in three studies, $A$ in none, so the mechanism-level claim, that outcomes depend on who performs the cognitive work, remains a strong, consistent, but not yet established candidate. Third, no finding in this evidence base licenses a population-scale claim in either direction, because the studies that identify effects are small and setting-specific, while the datasets with scale do not observe the outcome. That last statement is the gap this paper is about.

\section{What Current Measurement Misses}\label{sec:misses}

\subsection{The missing bridge}\label{sec:misses-bridge}

Table~\ref{tab:gap-matrix} is an illustrative evidence map of the studies reviewed in this paper; its inclusion rule is stated in the table note, and it claims coverage of this paper's evidence base, not of the literature. Two regularities organize it. Scale and outcome measurement rarely coincide: the deployment-telemetry datasets have millions of observations and no learning outcomes; the experimental rows have learning outcomes and, at most, about a thousand participants. The two rows with both, the adaptive-learning platform \citep{rismanchian2026faster} and the competitive-programming platform \citep{yao2026scaffold}, have them because assessment or competitive rating is built into the platform; the general-purpose assistants where student usage actually concentrates have no such instrumentation. And for general-purpose assistant use, none of the designs reviewed here links an individual's interaction traces to that same individual's unassisted retention and transfer at any meaningful scale, which is the specific missing object we call the bridge.

One research tradition comes close, and it is the model to import. Intelligent-tutoring-system and learning-analytics research routinely measures capability formation from interaction traces: dedicated platforms log every problem-solving step, model skills at fine grain, and support full experimental lifecycles, with one integrated platform reporting an estimated 147 studies across laboratory and classroom settings \citep{aleven2024ctat}. On such platforms, interaction content has been linked to learning directly, for example by combining dialog-act classification of tutoring chat with skill modeling to estimate which interaction types accompany higher learning rates \citep{borchers2024dialog}. The stock--formation gap is therefore not a claim that formation cannot be measured from traces; that is done routinely where the instrumentation was built for it. It is the observation that we find no equivalent instrumentation for the general-purpose assistants where student usage now concentrates, and the program in Section~\ref{sec:program} is in large part a proposal to carry this tradition's measurement discipline into that setting.

The education system's own instruments do not currently supply the bridge either. Assessment measures what students submit, not how capability was formed; a student who delegated an assignment and one who learned through dialogue can produce indistinguishable artifacts \citep{karamanis2026machines}, and even where responses differ, scores conflate what the student wrote with how the rater grades: in open-response analytics, modeling teachers' grading histories substantially improves grade prediction over response content alone \citep{borchers2026judgment}. AI-text detection is unreliable and inequitable in ways that preclude building measurement on it \citep{corbin2025}, and the automation of grading noted above removes human observation from precisely the point where formation would be noticed \citep{bent2025anthropic}. Privacy-preserving conversation classification demonstrates that interaction traces can be analyzed responsibly at scale \citep{tamkin2024clio}; what it currently surfaces is usage, not capability change.

\begin{table*}[t]
\centering
\caption{An illustrative evidence map: what the studies reviewed in this paper measure and what they miss.}
\label{tab:gap-matrix}
\small
\setlength{\tabcolsep}{4pt}
\begin{tabular}{p{2.5cm}p{1.0cm}p{1.6cm}p{1.2cm}p{1.0cm}c p{4.8cm}}
\toprule
\textbf{Study} & \textbf{Pop.} & \textbf{Scale} & \textbf{Design} & \textbf{Alloc.} & \textbf{Learn.} & \textbf{Headline finding} \\
\midrule
Deployment telemetry (four datasets) \citeyearpar{handa2025economic,anthropic2025student,bent2025anthropic,anthropic2026fluency} & Both & 9.8K--4M conv. & Tel. & --- & --- & usage patterns only \\
Fan et al.\ \mbox{\citeyearpar{fan2024}} & Form. & 117 students & RCT & Obs. & $\checkmark$ & performance up, learning flat \\
Shen \& Tamkin \mbox{\citeyearpar{shen2026skill}} & Bridge & 64 professionals & RCT & Obs. & $\checkmark$ & quiz deficit; pattern-dependent \\
Bastani et al.\ \mbox{\citeyearpar{bastani2025}} & Form. & $\approx$1K students & RCT & Yes & $\checkmark$ & deficit in base arm; not detected in scaffolded arm \\
Alsaiari et al.\ \mbox{\citeyearpar{alsaiari2026directive}} & Form. & 329 students & RCT & Yes & $\sim$ & hybrid feedback best; pure metacognitive worst \\
Kumar et al.\ \mbox{\citeyearpar{kumar2026empathy}} & Bridge & 968 adults & RCT & Yes & $\checkmark$ & gains under learner-works design \\
Contractor \& Reyes \mbox{\citeyearpar{contractor2026learning}} & Form. & 211 students & RCT & Obs. & $\checkmark$ & durable unaided gains; concentrated in explanation-seekers \\
Liu et al.\ \mbox{\citeyearpar{liu2026persistence}} & Mixed & 1{,}222 adults & RCT & Obs. & $\checkmark$ & unassisted performance harmed \\
Barcaui \mbox{\citeyearpar{barcaui2025crutch}} & Form. & 120 students & RCT & Obs. & $\checkmark$ & 45-day retention deficit \\
Rismanchian et al.\ \mbox{\citeyearpar{rismanchian2026faster}} & Form. & 3.2M interact. & Quasi & Obs. & $\checkmark^{\dagger}$ & proctored retention down, unproctored up \\
Yao \mbox{\citeyearpar{yao2026scaffold}} & Form. & CF submission histories & Tel. & Obs. & $\checkmark^{\dagger}$ & smaller gains for substitute-style users \\
StudyChat \mbox{\citeyearpar{mcnichols2026studychat}} & Form. & 181 students & Obs. & Obs. & $\checkmark$ & assisted--unassisted score gap; confounded (see \S\ref{sec:program-lessons}) \\
Liang et al.\ \mbox{\citeyearpar{liang2026systematic}} & Mixed & 56 studies & Rev. & --- & $\sim$ & internalization $>$ externalization \\
\bottomrule
\end{tabular}
\par\smallskip
\begin{minipage}{\textwidth}\footnotesize\raggedright
\emph{Note.} Pop.\ = population state (Stock = deploying formed capability; Form.\ = in formation; Bridge/Mixed/Both = spans both). Design: Tel.\ = observational platform or deployment telemetry; RCT = randomized trial; Quasi = quasi-experimental panel; Obs.\ = observational classroom data; Rev.\ = systematic review. Alloc.\ = whether the condition governing the allocation ($Z$) was randomized (Yes) or only realized usage ($A$) observed post hoc (Obs.). Learn.\ = unassisted learning outcome measured ($\checkmark$ = yes, $\sim$ = partial, --- = no). $^{\dagger}$Instrumented platforms with built-in assessment or competitive ratings; the general-purpose deployment-telemetry row has no outcome instrumentation. Inclusion rule: every deployment-scale usage dataset and every controlled or platform study measuring learning outcomes drawn on in this paper, plus the largest systematic review; studies were located by citation chasing and targeted search, so the map is illustrative of the reviewed evidence rather than systematic.
\end{minipage}
\end{table*}

\subsection{A descriptive illustration from public deployment data}\label{sec:misses-illustration}

\begin{figure}[ht]
\centering
\includegraphics[width=\columnwidth]{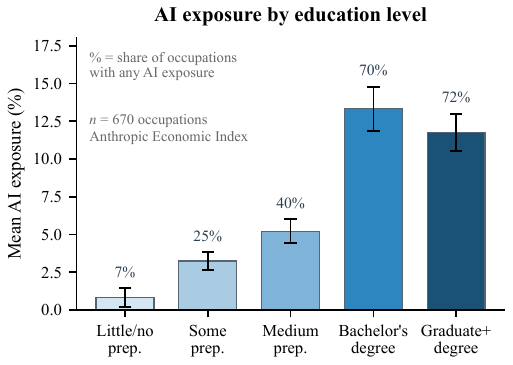}
\caption{Occupation-level (ecological) association: mean AI exposure by O*NET education zone, from public Anthropic Economic Index data \citep{handa2025economic} merged with O*NET education requirements ($n = 670$ occupations; percentages show the share of occupations with any observed exposure; error bars show $\pm 1$ standard error of the mean; Spearman $\rho = 0.425$). Descriptive only; see text for what this association can and cannot support.}
\label{fig:exposure}
\end{figure}

One reanalysis of public data illustrates which population the telemetry watches. We merged occupation-level AI exposure rates from the Economic Index with O*NET education requirements across 670 U.S.\ occupations (Figure~\ref{fig:exposure}). The association is monotonic and moderate-to-strong (Spearman $\rho = 0.425$, 95\% CI $[0.36, 0.49]$): occupations requiring a bachelor's degree or higher show mean exposure of 12--13\%, with 70--72\% registering any usage, while occupations requiring little formal preparation show mean exposure below 1\%. The association is descriptive, not causal: platform selection, task digitizability, and organizational policy all shape which occupations appear in assistant telemetry, and occupations are equally weighted. It is an ecological, occupation-level association: it shows that observed exposure concentrates in education-intensive occupations, not that the individuals behind the traces are in the stock state, since tenure, experience, and whether a user is currently learning a new domain are all unobserved. Read that way, it supports a deliberately narrow conclusion: deployment measurement is weighted toward skilled work, while the formation status of any given user, the quantity that matters for $\Delta K$, is exactly what such telemetry does not record. The formation side, whose outcomes the controlled studies show to be design-sensitive, is the side these instruments are least equipped to characterize.

\section{A Proposed Research Program}\label{sec:program}

The gap is closable with instruments that already exist. This section specifies the program: its design requirements, learned in part from our own pilot attempts; the bridge study at its center; the experimental arm that would test the design hypothesis; and the governance that must be built in rather than appended.

\subsection{Design requirements, including two lessons from our own pilots}\label{sec:program-lessons}

Two pilot analyses we conducted on public data shaped these requirements, chiefly by failing informatively.

First, we attempted to characterize the cognitive content of occupational tasks by classifying 2,644 O*NET task statements into taxonomy levels using their leading verbs. A two-coder audit on a stratified sample of 100 statements showed that humans code the construct reliably from full task text (inter-coder $\kappa = 0.80$) but agree with the verb heuristic only moderately ($\kappa = 0.49$; 68\% of disagreements between adjacent levels): the heuristic, not the construct, is the weak link. We therefore treat interaction-content classification as an instrument that must be validated before it is used, and the program below specifies full-text classification with human audit rather than surface heuristics.

Second, we reanalyzed the public StudyChat dataset \citep{mcnichols2026studychat}, in which 181 university students' assistant conversations coexist with course grades. Within student, AI-available assignment scores substantially exceed proctored no-AI exam scores, but assignments and exams differ in format, stakes, and timing as well as AI availability, so the gap cannot be attributed to AI; and contrasts between usage-profile groups were underpowered at this sample size for any plausible effect. The lesson is not that observational classroom data are useless; it is that the bridge requires topic-aligned, proctored, unassisted endpoints designed for the comparison, with power calculated in advance.

Combined with the evidence tiers, these lessons yield four requirements: (i)~interaction coding validated against human judgment, with classification uncertainty propagated into downstream estimates; (ii)~outcome measures that are unassisted, delayed, and include transfer, not only immediate performance; (iii)~experimental assignment of $Z$ wherever feasible, with realized allocation $A$ measured from traces rather than assumed, since Tier~2 shows usage patterns are self-selected; and (iv)~prospective, simulation-based multilevel power analysis that anticipates clustering, classifier error, and attenuation relative to trial-sized effects.

\subsection{The bridge study}\label{sec:program-bridge}

The core design links consented interaction traces to independent assessments within subject. Recruit students using a general-purpose assistant in ordinary coursework, with consent arrangements described below. Code each interaction, at the turn level, with a validated classifier audited against human-coded ground truth on stratified samples, on two axes. The first is the topic, to align interactions with assessment content. The second is the realized allocation $A$ in its profile reading: which of the component functions (planning, generation, monitoring, verification, correction) the learner, rather than the assistant, performs in the exchange; the two-dimensional structure of the revised Bloom's taxonomy \citep{anderson2001taxonomy}, which crosses cognitive process (Remember through Create) with knowledge type (factual through metacognitive), is a natural starting frame. The taxonomy's second dimension is the one current data miss most: existing usage classifications code, at best, the process level of what learners delegate, while the knowledge a learner brings to the exchange, the difference between an expert directing creation they can evaluate and a novice directing creation they cannot, is coded in none of the data sources reviewed here. Privacy-preserving classification of the kind already demonstrated at conversation scale makes such coding feasible without exposing conversation content \citep{tamkin2024clio}.

Link each student's usage profile to topic-aligned unassisted assessments: an immediate component, a delayed retention component, and a transfer component using novel problems from the same topic. Scoring such assessments at scale is itself tractable: LLM-scored open-ended items can reach conventional reliability thresholds with as few as fourteen items, using reliability estimation that accounts for the learning gains a pre-post design expects \citep{borchers2026reliability}. The within-subject structure controls stable individual differences, and baseline capability $K_0$ should be measured directly by pre-assessment rather than proxied. Identification is treated conservatively: where available, randomized defaults, randomized encouragement designs, or plausibly exogenous rollout variation could strengthen it; otherwise, associations between usage profiles and outcomes will be interpreted observationally.

Power should be planned by simulation rather than asserted. Students are clustered within courses and sites, and observational associations will be attenuated relative to the trial anchors ($d \approx 0.4$) by exposure heterogeneity and classifier error; attrition, outside-tool use, and model drift over a semester belong in the simulation as well. Planning for effects of $d = 0.1$--$0.2$ under a multilevel design is more credible than anchoring on trial-sized effects, and implies samples in the thousands with baseline covariates rather than hundreds. Datasets of this structure already exist in fragments (StudyChat demonstrates the trace-plus-grades linkage at small scale; the adaptive-learning platform demonstrates outcome instrumentation at large scale; and the tutoring-platform tradition of Section~\ref{sec:misses-bridge} has linked dialog-level interaction content to learning rates directly), so the barrier is assembly rather than invention, where assembly honestly means consent infrastructure, institutional data partnerships, and validated instruments rather than a download. The observational bridge cannot by itself establish causation, and it is not asked to: its role is to estimate, at deployment-relevant scale, whether the usage profiles the trials flag as risky are associated with lower unassisted retention and transfer, with what dose-response, and for whom.

\subsection{The experimental arm: varying the allocation}\label{sec:program-experiment}

The design hypothesis that Tier~1 motivates is specific and testable: holding the underlying model constant, assigned conditions that reserve the cognitive work for the learner (hints, feedback on attempts, evaluation of attempts) will produce equal or better $\Delta K$ than answer delivery, at some cost to immediate assisted performance, with the effect transmitted through the realized allocation $A$. The hypothesis is continuous with an older regularity in learning analytics, the doer effect: completing active practice predicts learning outcomes more strongly than passive consumption of material, an association recently extended beyond WEIRD populations and there moderated by prior educational attainment \citep{butler2025doer}. The experimental arm randomizes $Z$ directly: answer delivery versus hints versus feedback-on-attempt versus evaluation-of-attempt (the last including designs where the learner teaches or corrects the system \citep{mollick2023assigning}), crossed with learner expertise, since Tier~1 also shows that pure learner-side load can overwhelm novices \citep{alsaiari2026directive} and the frontier results show expertise changes what assistance does \citep{dell2023navigating}. Because assignment operates on $Z$ while the mechanism is hypothesized to run through $A$, the arm includes trace-based measurement of realized allocation as a manipulation check; that validation must show that coded $A$ varies within arms, since a classifier whose output merely recovers $Z$ measures the interface rather than the behavior. Mediation estimates of how much of the $Z$ effect $A$ transmits are exploratory on this design, because $A$ is post-treatment and measured with error; stronger identification would require designs that shift $A$ within $Z$, for example prompted-use variations under a fixed interface.

Deployed practice lowers this arm's cost: major assistants already ship learning modes built on the scaffolding principle \citep{anthropic2025learningmode,openai2025studymode}, and open deployments exist at scale \citep{buhler2026democratizing}, yet we found no learning-outcome evaluation of any deployed learning mode in the public record through mid-2026. Benchmarking work reinforces why that evaluation matters: prompted head-to-head against the instructional moves of a mature tutoring system, current LLMs only marginally reproduce its adaptivity, and the most instruction-following model tended toward exactly the overly direct, answer-oriented feedback that the unscaffolded trial arms instantiate \citep{borchers2025adaptivity}. Evaluating deployed learning modes is therefore a ready-made, high-value instance of this arm.

\subsection{Longitudinal extension and moderators}\label{sec:program-longitudinal}

Because formation compounds, single-course endpoints understate what matters. A longitudinal extension follows capability trajectories across courses or semesters, asking whether heavy delegation-style use in early courses predicts later performance in unassisted, cumulative settings, and whether effects fade, persist, or grow. Pre-registered moderators, beyond expertise, include task type, prior achievement, language background, and disability status, the latter two because both the benefits of access and the error rates of interaction classifiers are plausibly uneven across these groups.

\subsection{Governance by design}\label{sec:program-governance}

A program linking student conversations to assessment outcomes could itself become an instrument of surveillance or profiling, so its governance must be specified in advance, not retrofitted: participation requires meaningful consent with an opt-out carrying no academic penalty; measurements are controlled by a research function, not by instructors assigning grades or by disciplinary processes; retention is time-limited and labels auditable; classifications are never used for individual high-stakes decisions; and error rates are monitored for disparate impact, particularly for multilingual learners and learners with disabilities, for whom interaction-pattern classifiers are most likely to misread effort. Methods for auditing group-level model bias exist, but typical evaluation samples in educational data mining are underpowered for detecting realistic bias effects \citep{borchers2025abroca}, so the program's fairness audits must be powered deliberately rather than run as afterthoughts. Privacy-preserving classification \citep{tamkin2024clio} is a necessary component, not a sufficient one. We regard stakeholder co-design, with students, teachers, and assessment specialists, as a precondition for legitimate deployment of the program, and note that the present paper was developed without it.

\subsection{Exploratory: constructs the program may eventually need}\label{sec:program-constructs}

We flag, explicitly as a hypothesis space rather than a contribution, one further measurement question the program may encounter. The taxonomy the field uses to describe cognitive work \citep{anderson2001taxonomy} was built when no technology operated at its upper levels. If systems now produce competent output across those levels, the capacities that most distinguish human formation may sit above or beside the named ones: candidates include judgment about which problems are worth solving under genuine uncertainty, and the dispositional constructs the epistemic-cognition literature discusses under epistemic agency and identity, the self-directed initiation of verification and the stability of one's justificatory commitments \citep{greene2016handbook}. Each candidate would require discriminant validation against established constructs before use; sketch operationalizations exist (problem selection under an ill-posed brief; rate of self-initiated verification in think-aloud protocols; stability of commitments under counter-argument), and the think-aloud channel in particular is no longer prohibitive at scale: AI-transcribed think-alouds have been coded automatically for self-regulatory behaviors and related to moment-by-moment performance on tutoring platforms \citep{borchers2024thinkaloud}. Whether these constructs add predictive power over the standard taxonomy is itself a testable question the program could carry, and nothing else in the program depends on the answer.

\section{Implications and Limits}\label{sec:implications}

\subsection{What would count as evidence}\label{sec:implications-evidence}

The program is falsifiable in both directions, and it is worth stating in advance what results would mean. Evidence for harm would be: delegation-heavy usage profiles, those in which generation and verification sit mostly with the assistant, predicting lower unassisted retention and transfer, with dose-response, surviving selection controls, replicating across sites, and corroborated by the experimental arm, where assignment of $Z$ carries the causal weight. Evidence for benefit would be: profiles that retain the cognitive work, of which explanation-seeking use is the observed example, predicting equal or better unassisted outcomes alongside efficiency gains, as \citet{contractor2026learning} found in one proctored setting. The most likely outcome, on current evidence, is heterogeneity: effects that differ by allocation, expertise, and task, in which case the program's value is a map of where each design helps and harms, which is what procurement decisions, default-mode choices, and curriculum design actually need. In all three cases the decision-relevant sequence is the same: measure first, then set defaults; at present, defaults are being set first, at scale, and the measurement does not exist. Institutions listening to available data will hear the stock, whose productivity gains are measured, more clearly than the formation population, whose outcomes are not; that asymmetry, rather than any presumed direction of effect, is the policy reason the measurement should be built.

\subsection{Limitations}\label{sec:implications-limits}

This synthesis has limits that bound its claims. It is a motivated narrative synthesis, not a systematic review: studies were located by citation chasing and targeted search, the inclusion rule for Table~\ref{tab:gap-matrix} is stated in its note, and no risk-of-bias grading was performed; the map is illustrative rather than systematic. The assigned condition $Z$ was randomized in only three studies, and the realized allocation $A$ in none; elsewhere both are observational, and we have marked this accordingly throughout. The deployment data come disproportionately from a single vendor, partially triangulated by ChatGPT-based classroom data and platform telemetry; the experimental core does not depend on any vendor's data. Our own analyses are descriptive, and one pilot classifier failed its validation audit, which we report because the failure shaped the program's requirements (Section~\ref{sec:program-lessons}). The framing of education as the historical adjustment mechanism rests on US- and UK-centered economic history \citep{goldin2008,frey2019}, and its generalization is untested. The exploratory constructs of Section~\ref{sec:program-constructs} are hypotheses only. And the evidence base itself is young: most studies are small, several are preprints, and effect estimates carry publication-bias risk \citep{strohmaier2026llama}; conclusions here should move as it matures.

\subsection{Closing}\label{sec:closing}

The asymmetry of the stakes is why the program should not wait. When technological demand has outrun educational supply in the past, the costs took generations to repair \citep{goldin2008,frey2019}. If default interaction designs do erode formation, the harm will concentrate among learners with the least scaffolding and mentorship; adoption of learning-supportive tools already skews toward better-supported students in open deployments \citep{buhler2026democratizing}. If they do not, unfounded alarm carries its own costs in access withheld. Measurement is the inexpensive hedge against both errors, and Section~\ref{sec:implications-evidence} gave the reason it will not emerge on its own: the feedback channels institutions listen to are fed by the stock's measured gains, not by the formation side's unmeasured outcomes.

The question this paper formulates is narrow enough to answer and consequential enough to be worth answering: does sustained AI use, under the interaction designs people actually experience, change the formation of independent capability, in which direction, by how much, and for whom? The evidence reviewed here shows the question is real, shows it is currently unanswerable with deployed instruments, and shows that the core technical components of an answer already exist. Until the formation side is measured, decisions about defaults, deployment, and curriculum are being made without the one quantity that determines their long-run consequences. Education has been society's recovery mechanism through every previous technological disruption; whether it plays that role through this one should not rest on assumption. It is measurable, and this paper has tried to specify exactly how.

\bibliography{references}

\end{document}